\documentclass[a4paper,fleqn,usenatbib]{mnras}

\usepackage{amsmath}

\usepackage{times,txfonts}
\usepackage[T1]{fontenc}
\usepackage{aecompl}

\usepackage{graphicx}
\usepackage{color}
\usepackage{gensymb}
\usepackage{epstopdf}

\title[Identifying OHMs in ALFALFA]{Identifying OH Imposters in the ALFALFA Neutral Hydrogen Survey}

\author[K. A. Suess et al.]{
Katherine A. Suess,$^{1,2}$\thanks{E-mail: suess@berkeley.edu}
Jeremy Darling,$^{1}$
Martha P. Haynes$^{3}$
and Riccardo Giovanelli$^{3}$
\\
$^{1}$Center for Astrophysics and Space Astronomy, Department of Astrophysical and Planetary Sciences, University of Colorado, 389 UCB, \\Boulder, CO 80309-0389, USA\\
$^{2}$Department of Astronomy, University of California, Berkeley, 501 Campbell Hall, MC 3411, Berkeley, CA 94720-3411, USA\\
$^{3}$Cornell Center for Astrophysics and Planetary Science, 616A Space Sciences Building, Cornell University, Ithaca, NY 14853, USA
}

\date{Accepted 2016 March 16. Received 2016 March 09; in original form 2016 February 11.}

\pubyear{2015}

\begin{document}
\label{firstpage}
\pagerange{\pageref{firstpage}--\pageref{lastpage}}
\maketitle

\begin{abstract}
OH megamasers (OHMs) are rare, luminous molecular masers that are typically observed in (ultra) luminous infrared galaxies and serve as markers of major galaxy mergers. In blind emission line surveys such as the Arecibo Legacy Fast Arecibo L-Band Feed Array (ALFALFA) survey for neutral hydrogen (\ion{H}{i}), OHMs at z$\sim$0.2 can mimic z$\sim$0.05 \ion{H}{i} lines. We present the results of optical spectroscopy of ambiguous \ion{H}{i} detections in the ALFALFA 40\% data release detected by the Wide Field Infrared Survey Explorer (WISE) but with uncertain optical counterparts. The optical redshifts, obtained from observations at the Apache Point Observatory, revealed five new OHMs and identified 129 \ion{H}{i} optical counterparts. Sixty candidates remain ambiguous. The new OHMs are the first detected in a blind spectral line survey.

The number of OHMs in ALFALFA is consistent with predictions from the OH luminosity function. Additionally, the mid-infrared magnitudes and colors of the OHM host galaxies found in a blind survey do not seem to differ from those found in previous targeted surveys. This validates the methods used in previous IR-selected OHM surveys and indicates there is no previously unknown OHM-producing population at z$\sim$0.2. We also provide a method for future surveys to separate OH megamasers from 99\% of \ion{H}{i} line emitters without optical spectroscopy by using WISE infrared colors and magnitudes. Since the fraction of OHMs found in flux-limited \ion{H}{i} surveys is expected to increase with the survey's redshift, this selection method can be applied to future flux-limited high-redshift hydrogen surveys.
\end{abstract}




\section{Introduction}

The first hydroxyl megamaser (OHM) was discovered by \citet{baanArp} in Arp 220. The luminosity of the galaxy's OH emission line exceeded $10^3 L_\odot$ (orders of magnitude more luminous than known galactic OH masers), leading to the rise of the term `megamaser' to describe masers with isotropic line luminosities in the range $10^{1-4} L_\odot$. Early OHM surveys such as \citet{baan85} focused on galaxies with bright radio continuum and found $<$20 OHMs. After the launch of IRAS, surveys such as \citet{staveley} and \citet{norris} used IR properties to select ULIRG-like candidates with flat far-IR spectral indices and steep mid-IR spectral indices; these surveys discovered $\sim$30 OHMs. \citet{oh1}, \citet{oh2}, and \citet{oh3} (hereafter the Arecibo Megamaser Survey) carried out a deep survey using IR selection criteria that roughly doubled the number of known OHMs. There are $\sim$110 known OHMs up to $z=0.264$, most of which are listed in \citet{oh3}. 

All known OHMs are found in (ultra) luminous infrared galaxies ([U]LIRGs), extreme starburst galaxies that are almost exclusively the products of major galaxy mergers \citep{clements}. Merger phase is correlated with the far infrared (FIR) luminosity of ULIRGs, and the OHM fraction in ULIRGs is a strong function of the FIR luminosity \citep{baan}; this suggests that the presence of an OHM in a ULIRG indicates the phase of the merger. OHMs are also associated with high dense molecular gas fractions \citep{gasfrac}, further indicating their relation to merger phase. Because OHMs are observable at large distances, they could provide a useful tracer of the galaxy merger rate as a function of redshift. Zeeman splitting of the OH line has also been observed in several OHMs \citep{zeeman}, allowing for direct measurement of magnetic fields in star-forming regions. 

While OHMs are interesting in their own right, they can also contaminate in blind emission line surveys for neutral hydrogen (\ion{H}{i}). If $\nu_{\ion{H}{i}, 0}/(1+z_{\ion{H}{i}}) = \nu_{OH, 0}/(1+ z_{OH})$, OH and \ion{H}{i} lines appear at the same frequency and OHMs can be mis-identified as \ion{H}{i} sources. While the fraction of OHMs in low-redshift \ion{H}{i} surveys is small, this fraction is expected to increase with redshift and reach 50\% by $z=1$ \citep{briggs}. It is therefore necessary to develop a method to separate OH and \ion{H}{i} lines in blind spectral line surveys before the advent of high-redshift \ion{H}{i} surveys \citep[e.g. ASKAP,][]{askap}.

In this work, we present the first OHMs detected in a blind spectral line survey. Five new OHMs at $0.167\le z_{OH} \le 0.244$ were detected in the 40\% data release of the Arecibo Legacy Fast Arecibo L-Band Feed Array Survey \citep[ALFALFA,][]{alpha40}, a blind emission line survey for \ion{H}{i} at $z \le 0.06$. After confirming that the number and IR properties of these OHMs match empirical predictions, we develop a method to separate OH from \ion{H}{i} lines without the use of optical spectroscopy.

Throughout this work, we assume a $\Lambda$CDM cosmology with $\Omega_m = 0.29,\ \Omega_\Lambda = 0.71$, and $H_0 = 70$km s$^{-1}$ Mpc$^{-1}$ \citep{wmap}.

\section{Observations and Data Reduction}

We identified a total of 194 objects as potential OHMs by selecting objects from the ALFALFA 40\% data release \citep{alpha40} that had no optical counterpart velocity or an optical velocity that differed from the HI velocity by more than 300 km s$^{-1}$.  We rejected large spiral galaxies that are clearly HI emitters. When more than one possible optical counterpart fell within the ALFALFA position ellipse, we selected the closest WISE mid-IR counterpart detected at 22 $\mu$m for optical spectroscopy.

We observed candidate OHMs at the Apache Point Observatory over 15 sessions between December 2011 and March 2013. Observations were made using the Dual Imaging Spectrograph at the Apache Point Observatory (APO) 3.5 m telescope with the B400/R300 grating and a 1.5" spectroscopic slit. This setup has a dichroic wavelength of $\sim$5350~\AA\ and a resolving power of R$\sim$3250 on the red side and R$\sim$2400 on the blue side. Most targets had 2-4 observed frames, for a total exposure time of 5-20 minutes for typical objects and 1-2 hours for the faintest targets. Calibration images included bias frames, quartz lamp flat field frames, and spectra of helium, neon, and argon arc lamps for wavelength calibration.

Data reduction followed standard procedures for the IRAF `longslit' package. Upper bounds for the pixel-\AA\ axis transformation uncertainty were 4~km~s$^{-1}$ on the red side and 21~km~s$^{-1}$ on the blue side. We adjusted the wavelength solution to heliocentric velocity using `rvcorrect' and `dopcor'; this process added around 1~km~s$^{-1}$ of uncertainty in the wavelength solutions. Upper bounds for the wavelength calibration uncertainty were 125~km~s$^{-1}$ on the blue side and 92 km~s$^{-1}$ on the red side. After calibration, we aligned and median stacked exposures of the same science target. 

We made line measurements using the IRAF `splot' task. The red side of the spectrum usually showed H$\alpha$ (6563~\AA) bracketed by two [NII] lines (6549 and 6583 \AA), with a [SII] doublet (6717 and 6731 \AA) on the redward side. On the blue side, the most common lines were H$\beta$ (4861 \AA) and two [OIII] lines (5007 and 4959 \AA). We also commonly observed other lines in the Balmer series of hydrogen as well as the 3727 \AA\ [OII] line on the blue side. Most objects had at least 5-7 observed optical lines, with 9 lines common. Only in a few cases were fewer than 5 lines observed. We measured RMS noise for each target in clean regions of the spectrum, away from astronomical sources, night sky lines, and cosmic rays. We calculated final redshifts using an error-weighted average of the individual line measurements. The typical final uncertainty in redshift was $2~\times 10^{-6}$, or 0.6~km~s$^{-1}$. The maximum centroid uncertainty observed was $5.4~\times 10^{-5}$, or 16~km~s$^{-1}$. 

Typical and maximum values for all factors contributing to the the final uncertainty in measured velocities are: instrumental uncertainty, 112 km~s$^{-1}$ typical and 125 km~s$^{-1}$ maximum; wavelength calibration, 13 km~s$^{-1}$ typical and 21~km~s$^{-1}$ maximum; heliocentric calibration, 1~km~s$^{-1}$ maximum; and line centroid uncertainty, 1~km~s$^{-1}$ typical and 16~km~s$^{-1}$ maximum. Adding these uncertainties in quadrature, we arrive at 113~km~s$^{-1}$ typical uncertainty and 128~km~s$^{-1}$ maximum uncertainty. We thus conservatively adopt 130~km~s$^{-1}$ as the uncertainty on all optical recession velocity measurements.


\section{Results}
For each observation, we could make one of three determinations. The first, and most common, was that the velocity of the observed object matched (within uncertainty) the velocity listed in the ALFALFA catalog. It was also possible for the object's velocity to match the OH velocity, found by recalculating the velocity in the ALFALFA catalog using an OH rest frequency instead of an \ion{H}{i} rest frequency:
\begin{equation}
z_{OH} = \frac{\nu _{OH, 0}}{\nu _{\ion{H}{i}, 0}} (1+ z_{\ion{H}{i}}) -1,
\label{zOH}
\end{equation}
where $z_i = v_i / c$ and we adopt a frequency of 1667.35903 MHz for OH and 1420.405752 MHz for \ion{H}{i}. These objects are OHMs. The third possible outcome was ambiguous, where the observed velocity matched neither the \ion{H}{i} nor the OH velocity. These ambiguous cases are observations of the incorrect optical counterpart or of a false positive ALFALFA detection.

\subsection{OH Detections}
\begin{figure*}
	\centering
	\includegraphics[width=\textwidth]{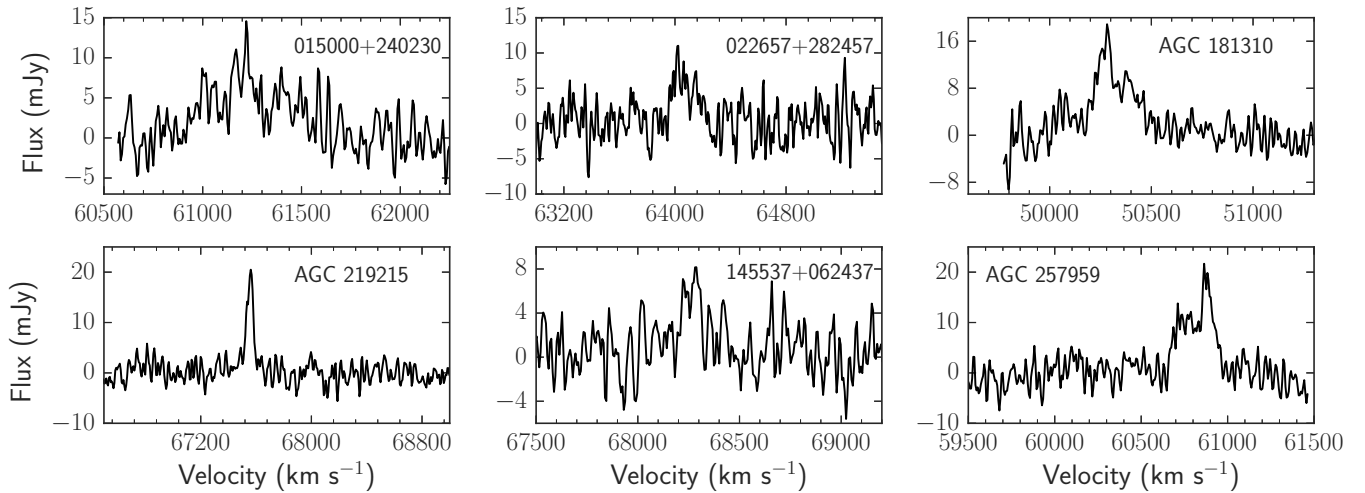}
	\caption{ALFALFA spectra of the six OH megamasers confirmed with APO observations. The velocity axis indicates the heliocentric optical velocity of the OH 1667 MHz line.}
	\label{fig:spectra}
\end{figure*}

We identified five previously undiscovered OHMs through APO observations. Additionally, \citet{alpha40} identified one previously discovered OHM \citep[AGC 181310, IRAS 08201+2801, discovered by][]{oh2} in the ALFALFA $\alpha$.40 release. These six objects are the only known OHMs in the ALFALFA $\alpha$.40 database. Extrapolating to the full sample size indicates ALFALFA will detect on the order of 15 total new OHMs, slightly lower than \citet{ALFALFAprop} predictions that the survey `should detect several additional dozen OHMs.' Table \ref{OHmatch} lists the optical, OH, and \ion{H}{i} velocities for the six ALFALFA OHMs as well as the line peak and width from ALFALFA \citep{alpha40}. ALFALFA extracts line peaks and widths using a matched-filtering approach described in \citet{saintonge}; the templates used are Hermite functions. Table \ref{OHprop} lists the infrared properties of the OHMs. Figure \ref{fig:spectra} shows the spectra of the six OHMs.

\begin{table*}
\small
\caption{OH Megamasers detected in the ALFALFA 40\% survey. Optical velocities were obtained from APO observations; all uncertainties are 130 km s$^{-1}$. \ion{H}{i} velocities are taken from the ALFALFA $\alpha$.40 data release \citep{alpha40}, with uncertainties in parentheses. OH velocities are calculated from the \ion{H}{i} velocities using Equation \ref{zOH}, with uncertainties in parentheses. The bolded OHM was discovered in the $\alpha$.40 data release by \citet{alpha40}.}
\label{OHmatch}
	\begin{tabular}{ccccccc}
	\hline
	{Object Name} & {Position (J2000)} & \multicolumn{1}{|c|}{\parbox{2.2cm}{\centering{Optical Velocity (km s$^{-1}$)}}} & \multicolumn{1}{|c|}{\parbox{2.1cm}{\centering{\ion{H}{i} Velocity (km~s$^{-1}$)}}} & \multicolumn{1}{|c|}{\parbox{2.1cm}{\centering{OH Velocity (km~s$^{-1}$)}}} & \multicolumn{1}{|c|}{\parbox{2.1cm}{\centering{Line Peak (mJy)}}} &  \multicolumn{1}{|c|}{\parbox{2.1cm}{\centering{Line Width  (km~s$^{-1}$)}}} \\ \hline
	015001+240236 & 015001.57+240235.8 & 61268 & 7775(32) & 61249(38) & 9(2) & 468(64)\\
	022657+282457 & 022657.65+282457.5 & 64397 & 10185(11) & 64078(13) &11(3) & 158(22)\\
	\textbf{AGC 181310} &  \textbf{082312.7+275138} & \textbf{50365} & \textbf{-1551(13)} & \textbf{50302(15)} & \textbf{19(2)} & \textbf{193(26)} \\
	AGC 219215 & 111125.06+052046.0 & 67517 & 13148(2) & 67556(2) & 21(2) & 45(4)\\
	145537+062437 & 145537.39+062437.4 & 68960 & 13749(5) & 68262(6) & 8(2) & 83(9)\\
	AGC 257959 & 155537.94+143905.6 & 61028 & 7393(2) & 60801(2) & 20(2) & 206(5) \\ \hline
	\end{tabular}
\end{table*}

\begin{table*}
\small
\caption{Luminosity and IR properties of OH Megamasers detected in the ALFALFA 40\% survey. OH luminosity is calculated from the ALFALFA detection, and the far infrared (FIR) luminosity is derived from the 60 and 100 $\mu$m IRAS fluxes according to the prescription in \citet{FIRL}. The notation [22] refers to the Vega magnitude measured at 22 $\mu$m.}
\label{OHprop}
	\begin{tabular}{ccccccccc}
	\hline 
	{{Object Name}} & {$\log(\frac{L_{OH}}{L_{\odot}})$} & WISE [3.4] & WISE [4.6] & WISE [12] & WISE [22] & {IRAS $f_{60 \mu m}$} (Jy) & {IRAS $f_{100 \mu m}$} (Jy) & {$\log(\frac{L_{FIR}}{L_\odot})$} \\ \hline 

	015001+240236 & 3.61 & 14.23 & 13.61 & 9.82 & 7.47 & 0.301 & 1.064 &  11.87\\
	022657+282457 & 3.29 & 13.39 & 12.55 & 9.43 & 7.29 &  -- & -- & --  \\
	AGC 181310 &  3.33 & 14.20 & 13.22 & 8.33 & 5.02 &1.171 & 1.430 & 11.97 \\
	AGC 219215 & 3.26 & 15.15 & 13.99 & 9.94 & 7.59 & -- & -- & -- \\
	145537+062437 & 3.06 & 14.60 & 13.72 & 9.54 & 6.60 & 0.470 & 1.388 & 12.14 \\
	AGC 257959 & 3.60 & 14.91 & 13.54 & 9.39 & 7.09 & 0.743 & 1.194 & 12.10 \\
	\hline
	\end{tabular}
\end{table*}

\subsection{H {\normalsize \textbf{I}} confirmations}

We confirmed 129 \ion{H}{i} sources with APO observations. For each, we noted the exact location of the optical counterpart-- these occasionally differed slightly from the ALFALFA location due to the survey's large beam. We used SDSS DR9 in conjunction with telescope pointing images taken during APO observations to confirm the J2000 coordinates of the \ion{H}{i} optical counterpart, and APO telescope pointing images to confirm the positions of the few objects that did not fall within the Sloan sky coverage. In Table~\ref{HImatch}, we list all identified \ion{H}{i} sources with their name (6-numeral strings represent AGC name), optical position, optical velocity, and ALFALFA velocity.
\begin{table*}
\footnotesize
\caption{Optical counterpart redshifts of \ion{H}{i}-emitting galaxies. Six-digit object names are AGC names. \ion{H}{i} velocities are from \citet{alpha40}, with uncertainties in parentheses.}
\begin{tabular}{cccc}
\hline {Object Name} & {Optical Counterpart Position (J2000)} & {Optical Velocity$^1$ (km s$^{-1}$)} & {\ion{H}{i} Velocity (km s$^{-1}$)}
\\ \hline
102733 & 	000129.98+311402.9 & 12660 &  12581(6) \\
100783	 & 000347.49+312037.7 & 	5065 & 	5011(4) \\
102902 & 	000948.86+284123.8 & 	10688 & 	10560(4) \\
107 & 	001138.10+275652.8  & 	7488	 & 7437(3) \\
102643 & 	002136.21+252931.1 & 	6960 & 	7042(13) \\
102644 & 	002251.57+254720.2 & 	7188	 & 7018(10) \\
HI003023+251839 & 003023.94+251903.8 &  	7424 & 7295(7) \\
100291 & 	003212.12+312459.5 & 	6232 & 	6140(30) \\
102301 & 	004342.83+255150.1 & 	5167 & 	5180(5) \\
102935 & 004825.32+284535.4 & 4475 & 4400(10) \\
101685 & 	010034.02+270552.7 & 	11216 & 	11040(4) \\
113892 & 	010105.79+310419.4  & 	6677 & 	6714(3) \\
748808 & 	010908.52+141358.4 & 	9551 & 	9551(3) \\
113924 & 	012626.02+310032.5 & 	13646	 & 13470(3) \\
114047 & 013326.39+285623.4 & 	7801 & 	7700(8) \\
113941 & 	014140.52+312946.6 & 	10975	 & 10741(2) \\
114121 & 	014636.92+144131.0 & 	7462	 & 7485(2) \\
748822 & 	015224.00+154101.6 & 	13036 & 	12866(8) \\
113956 & 	015327.57+305349.9 & 	11353 & 	11205(2) \\
113964 & 	015521.61+313730.4 & 	4634	 & 4486(3) \\
123118 & 	020149.86+292647.7 & 	16988 & 	16909(16) \\
122960 & 	020426.95+310734.6 & 	4907 & 	4787(4) \\
HI020624.2+160545 & 020626.40+160538.1 & 	5308 & 	5346(4) \\
748838 & 	020948.79+153345.1 & 	10934	 & 10774(4) \\
748840 & 021022.44+144415.7 & 	17764 & 	17578(11) \\
122141 & 	021133.13+141419.4 & 	3807 & 	3797(3) \\
123103 & 	021145.35+311130.2 & 	4740 & 	4819(5) \\
121499 & 	021228.93+291057.9 & 	9996 & 	9920(5) \\
122979 & 	021513.92+310533.3 & 	5007 & 	4968(11) \\
122855 & 	021847.24+145042.8 & 	3996 & 	3907(2) \\
122988 & 	021954.56+295050.3 & 	11230 & 	11080(12) \\
122194 & 	022139.74+280307.2 & 	10783 & 	10642(4) \\
120193 & 	022255.51+251827.4 & 	4665	 & 4584(10) \\
123143 & 	022312.08+282722.9 & 	10321 & 	10295(3) \\
122859 & 	022451.18+161039.3 & 	8291 & 	8189(3) \\
123005$^2$ & 	022542.30+313722.9 & 	5035 & 	5159(20) \\
120240 & 	022555.82+245125.0 & 	10890	 & 10799(8) \\
121216 & 	022558.74+271613.5 & 	10345 & 	10171(42) \\
123158 & 	023006.35+284027.2 & 	11120 & 	10958(2) \\
122883 & 	023141.96+241720.6 & 	5338 & 	5409(8) \\
122214 & 	023233.70+275626.8 & 	4777 & 	4657(7) \\
123163 & 	023245.42+283318.5 & 	10678 & 	10660(7) \\
122215 & 	023328.44+271140.9 & 	5471 & 	5343(7) \\
120342 & 	023346.91+301121.5 & 	10339 & 	10253(5) \\
2155	 & 024005.68+142233.3	 & 13893 & 	13911(3) \\
122421 & 	024131.69+263742.9 &  	1566	 & 1586(3) \\
748875 & 	024604.91+143916.1 & 	7580	 & 7541(13) \\
122850 & 	024639.50+150856.6 & 	7828 & 	7791(3) \\
122857 & 	024703.76+145052.0 & 	7761 & 	7598(2) \\
120529 & 	024853.52+281624.8 & 	5432 & 	5424(6) \\
748888 & 	024922.00+150223.7 & 	8824 & 	8819(32) \\
749015 & 	025303.78+143739.3  & 	9829 & 	9891(33) \\
122809 & 	025905.53+271055.4 & 	10853	 & 10842(10) \\
122810 & 025929.19+255351.4 & 10358 & 10402(5) \\
123065 & 	025959.37+305912.8 & 	5924 & 	5971(2) \\
748916 & 	030647.43+143633.0 & 	10105 & 	10014(4) \\

\label{HImatch}
\end{tabular}

\medskip
{$^1$}{The uncertainty on optical velocities is 130 km s$^{-1}$ for all objects.}

$^2${This spectrum showed both emission and absorption lines. The emission lines were more prevalent, but the offset absorption lines clearly showed in the blue half of the spectrum. Absorption was especially evident in H$\beta$ and blueward Balmer series lines, where broad emission lines were nearly divided in two by the offset absorption. The emission velocity occurred at 5035 km s$^{-1}$, a match for the ALFALFA \ion{H}{i} line; the absorption lines were offset to 5770 km s$^{-1}$. The velocity listed in Table~\ref{HImatch} is the emission velocity, which matches the ALFALFA \ion{H}{i} velocity. The galaxy exhibits disturbed morphology in SDSS, and the absorption and emission lines may be offset because they come from different nuclei.}
\end{table*}

\begin{table*}
\contcaption{Optical Counterpart Redshifts of \ion{H}{i}-Emitting Galaxies.}
\footnotesize
\begin{tabular}{cccc}
\hline {Object Name} & {Optical Counterpart Position (J2000)} & {Optical Velocity (km s$^{-1}$)} & {\ion{H}{i} Velocity (km s$^{-1}$)}
\\ \hline
748918 & 	030715.30+151745.7 & 	5687 & 	5659(8) \\
174684 & 	073019.16+060634.9 & 	8502 & 	8504(2) \\
174697 & 	073532.66+062646.3 & 	9719 & 	9755(3) \\
174481 & 	073602.49+133216.4 & 	4740 & 	4770(4) \\
174491 & 073732.29+125218.4 & 	13970 & 	13973(5) \\
170347 & 	074035.08+260806.1 & 	8394 & 	8383(7) \\
182739 & 	080247.11+244617.0 & 	12463 & 	12379(25) \\
188943 & 	080520.82+055706.6 & 	9055	 & 9100(34) \\
180967 & 	081425.34+042032.9 & 	10117 & 	10275(6) \\
749273 & 081709.90+263354.2 & 	5817	 & 5838(4) \\
182496 & 	082626.06+044839.2 & 	8531 & 	8525(2) \\
183495 & 	082907.22+275655.0 & 	12525 & 	12568(11) \\
184464 & 	085402.71+275730.5 & 	8094	 & 8006(6) \\
749210 & 	113201.37+272451.2 & 	15203 & 	15050(3) \\
215140 & 	114201.25+134155.5 & 	4435 & 	4259(4) \\
HI122922+042247$^3$ & 122922.81+042246.1 & 	4999 & 	5009(4) \\
221030 & 	124835.53+090732.3 & 	7711 & 	7558(4) \\
238878 & 	130212.70+110034.0 & 	13636 & 	13689(36) \\
230239 & 131928.51+143439.4 & 	6787	 & 6705(6) \\
238831$^4$ & 	132102.19+260833.4 & 	17192 & 	17081(4) \\
749554 & 132537.83+244712.5 & 	10094 & 	10076(4) \\
233819 & 	135106.13+082038.9 & 	6971 & 	6915(13) \\
241309 & 	140437.88+152832.0 & 	7904	 & 7786(9) \\
240736 & 	144934.54+111453.1 & 	16212 & 	16426(4) \\
249263$^5$ & 145233.39+060115.3  & East: 14196 & 	14327(3) \\
 & 145235.88+060128.6 & South: 14452 & \\
248894 & 	145949.34+152421.3 & 	13621 & 	13528(7) \\
258105 & 	150157.30+091118.1 & 	8988 & 	8895(17) \\
258530 & 	150834.82+265155.4 & 	17523 & 	17513(29) \\
257884 & 	151333.16+121211.0 & 	16638 & 	16696(4) \\
727058 & 	151956.28+253618.2 & 	9789	 & 9695(30) \\
257934 & 	154349.18+143856.7 & 	10440 & 	10363(9) \\
258337 & 	154555.99+043249.7 & 	6499	 & 6440(1) \\
257961 & 	155637.52+160224.1 & 	4617 & 	4538(4) \\
268208 & 	162934.08+040227.3 & 	16315 & 	16255(4) \\
268223 & 	162942.53+055505.3  & 	9912 & 	9866(3) \\
748649 & 	214534.16+135511.8 & 	8774 & 	8788(39) \\
310185 & 	215016.37+155234.8 & 	7597 & 	7520(3) \\
310204 & 	215252.88+153418.5 & 	13363	 & 13169(3) \\
748661 & 	215352.36+160637.2 & 	7840 & 	7691(3) \\
321219 & 	220954.17+263157.1 & 	11443	 & 11355(7) \\
321410 & 	221133.62+305412.1 & 	4934	 & 4805(4) \\
321209 & 	222121.00+275033.3 & 	12665 & 	12706(13) \\
320185 & 	222355.67+151447.3 & 	7398 & 	7318(3) \\
321344 & 	223531.15+251040.0 & 	12220 & 	12108(3) \\
321284 & 223827.40+255502.7 & 	8761 & 	8593(11) \\
748716 & 	224105.61+154923.9 & 	1935	 & 1936(3) \\
321487 & 	224230.83+293229.8 & 	7475 & 	7314(5) \\
320379 & 	224833.94+243205.0 & 	12441 & 	12319(12) \\
321440$^6$ & 	225030.19+315112.5 & 	6418 & 	6462(1) \\
321453 & 225720.88+315316.0 & 	6682 & 	6660(5) \\
332908 & 	230543.41+271245.7 & 	7457 & 	7406(3) \\
332417 &  230636.70+141014.7 & 	12150 & 	11962(6) \\
333370 & 	231422.07+311503.4 & 	7286	 & 7235(9) \\
HI231551+253430 & 231551.33+253428.5 & 9900 & 9829(16) \\
333634 & 	231643.23+244148.4 & 	17193 & 	17155(141) \\
333525 & 	232017.84+290859.5 & 	6117 & 	5956(17) \\
\end{tabular}

\medskip 
$^3${This object was identified as a clear \ion{H}{i} detection; in SDSS it appears to be a blue compact dwarf galaxy.}

$^4${The optical counterpart for this object was confirmed to be the blue extended object to the west of the central ALFALFA coordinates.}

$^5${Two objects were observed within the ALFALFA beam radius. One observed galaxy was to the east of the ALFALFA target coordinates, and one to the south. Both galaxies matched the \ion{H}{i} velocity, so the coordinates and measured velocities for both are included in this table.}

$^6${Most observed objects have at least five identifying spectral lines. For this object, only H$\alpha$ was identified. Despite the scarcity of observable optical lines in this spectra, we are confident of the \ion{H}{i} confirmation. The singular line is bright, shows extension along the spatial axis of the spectrum, and it is very close to the expected \ion{H}{i} redshift. In visible light, the object appears to be a low surface brightness galaxy.}
\end{table*}

\begin{table*}
\contcaption{Optical Counterpart Redshifts of \ion{H}{i}-Emitting Galaxies.}
\footnotesize
\begin{tabular}{cccc}
\hline {Object Name} & {Optical Counterpart Position (J2000)} & {Optical Velocity (km s$^{-1}$)} & {\ion{H}{i} Velocity (km s$^{-1}$)}
\\ \hline
333331 & 	232510.17+245047.8 & 	9766 & 	9748(4) \\
333286 & 	232551.71+253820.8 & 	8732 & 	8582(21) \\
333538 & 	232637.07+294125.1 & 	6837 & 	6788(3) \\
333392 & 	232708.90+302417.9 & 	4541 & 	4521(4) \\
333460 & 	232854.66+310459.5 & 	13637 & 	13641(7) \\
333398 & 	232947.28+301524.8 & 	8393	 & 8269(8) \\
331198 & 	233211.07+285731.5 & 	5615 & 	5512(23) \\
12658 & 	233244.79+310649.4 & 	9615 & 	9502(2) \\
331305 & 	234324.28+265457.3 & 	8252 & 	8189(5) \\
333419 & 	234411.82+314557.8  & 	9395 & 	9319(5) \\
333566 & 	234557.53+290958.4 & 	9784 & 	9694(40) \\
333232 & 	234628.30+274423.6 & 	8106 & 	8092(2) \\
333205 & 	234629.34+274131.6  & 	16815	 & 16873(12) \\
331380 & 	235347.85+253529.7 & 	11423	 & 11510(10) \\
333220 & 	235529.44+275902.3 & 	9191 & 	9015(6) \\
333436 & 	235648.68+302422.5 & 	9455 & 	9331(11) \\
333239 & 	235916.84+274521.4 & 	14615 & 	14586(5) \\
\end{tabular}
\end{table*}

\subsection{Ambiguous Optical Counterparts}
Sixty objects remained ambiguous after APO observations: velocities for these objects matched neither \ion{H}{i} nor OH velocities, or no optical lines were detected. Many of the observed OHM candidates were not high signal-to-noise ALFALFA detections, and the majority of the ambiguous optical counterparts are likely false positive ALFALFA detections. When multiple possible optical counterparts appeared within the ALFALFA beam radius, we made observations of the object brightest in the WISE 22 $\mu$m band; some of the ambiguous optical counterparts likely correspond to observations of a different object than the ALFALFA emitter. Several ambiguous objects had no observable optical lines. Table~\ref{ambig} lists ambiguous optical counterparts.
\begin{table*}
\footnotesize
\caption{Ambiguous optical counterparts. Six-digit object names are AGC names. \ion{H}{i} velocities are from \citet{alpha40}, with uncertainties in parentheses. The designation `neither' indicates that the observed velocity matches neither \ion{H}{i} nor OH, while `no lines' designates objects for which no optical lines were observable and no velocity determination could be made. `ML' indicates that the ALFALFA \ion{H}{i} detection is a marginal line, likely not a real detections. `UL' indicates that the ALFALFA line is uncertain, and could be \ion{H}{i} or another line.}
\begin{tabular}{cccccc}
\hline {Object Name} & \multicolumn{1}{|c|}{\parbox{2.1cm}{\centering Observed Position (J2000)}} & {Optical Velocity$^1$ (km s$^{-1}$)} & {\ion{H}{i} Velocity (km s$^{-1}$)} & {OH Velocity (km s$^{-1}$)} & {Designation}
\\ \hline

HI000335.7+253214 & 000336.02+253204.0 & - & -1319(7) & 50573(8) & No Lines \\
102624 & 000903.33+252750.4 & 31506 & 1393(5) & 53757(6) & Neither, UL \\
HI002048+294651$^2$ & 002049.69+294651.5 & - & 6799(4) & 60103(5) & Wrong Pointing \\
HI002957+305739$^3$ & 002957.16+305744.3 & 22618 & -596(5) & 51423(6) & Neither \\
102983 & 003338.74+284526.0 & - & 3607(9) & 56356(11) & No Lines \\
HI005058+284800 & \multicolumn{1}{|c|}{\parbox{2.5cm}{A:005058.30+284800.0; B:005059.40+284803.6}} & A: 87656; B: 87233 & 1255(20) & 53595(23) & Neither, UL \\
102942 & 005355.45+290715.8 & 10936 & 11585(5) & 65721(6) & Neither \\
HI005555+294810 & 005555.11+294836.2 & 107964 & 1038(9) & 53341(11) & Neither, UL \\
102820$^4$ & 005626.11+305408.2 & 5320 & 4776(21) & 57729(25) & Neither \\
HI011145+290458 & 011142.86+290510.9 & A: 28224; B: 28710 & 16654(18) & 71672(21) & Neither, ML \\
HI011200+274341 & 011200.29+274342.0 & 33390 & 14232(3) & 219192(4) & Neither, ML \\
HI012215+284810 & 012217.19+284753.0 & 4321 & 2161(4) & 54659(5) & Neither \\
114080 & 013310.10+284520.3 & 193966 & 1865(29) & 54311(34) & Neither, ML, AGN \\
113863 & 014339.23+260036.2 & 39842 & 3932(20) & 56738(23) & Neither, ML \\
HI015722.6+144843$^5$ & 015725.33+144813.9 & - & 7589(4) & 61030(5) & Neither \\
113868 & 015838.90+250726.6 & 30843 & 1922(5) & 54378(6) & Neither, ML \\
HI020827.4+154646 & 020828.15+154639.3 & - & 4701(4) & 57640(5) & No Lines \\
HI021034.5+253405 & 021033.59+253327.5 & - & 17647(5) & 72837(6) & No Lines \\
121286 & 021646.75+291236.2 & 18988 & 13052(10) & 67443(12) & Neither, UL \\
HI022701.3+245402 & 022702.92+245421.2 & 64392 & 14966(20) & 69690(23) & Neither, UL \\
122433 & 023126.73+255653.9 & 15836 & 14223(14) & 68818(16) & Neither \\
HI073435.3+083500 & 073436.24+083500.6 & 47004 & 168(5) & 52319(6) & Neither, UL \\
174555 & 074945.10+134554.5 & 29544 & 2416(7) & 54958(8) & Neither, UL \\
HI080838.6+053210 & 080840.87+053140.4 & - & 9187(14) & 62906(16) & No Lines \\
189051 & 085016.90+271219.0 & 29971 & 4515(4) & 57422(5) & Neither, UL \\
219219 & 110416.13+045719.9 & 66771 & 7113(6) & 60472(7) & Neither \\
219220 & 111301.97+040305.7 & 45550 & 6683(8) & 59967(9) & Neither \\
215280 & 111316.20+152431.8 & - & 1479(2) & 53858(2) & No Lines \\
215238 & 111519.91+114053.6 & 50092 & 3053(4) & 55706(5) & Neither, UL \\
219222 & 111553.88+042227.9 & 24701 & 6218(28) & 81118(33) & Neither, ML \\
HI113900.7+102250 & 113858.83+102222.5 & 73121 & 5118(5) & 58130(6) & Neither, ML \\
HI115119.4+274818 & 115121.03+274827.0 & 26848 & 14191(5) & 68780(6) & Neither, UL \\
HI124540+070337 & 124543.33+070329.2 & - & -624(4) & 51390(5) & No Lines \\ 
HI130227+135524 & 130226.79+135548.7 & 83292 & 13344(1) & 67786(1) & Neither \\
HI134330.3+111234$^6$ & 134332.00+111220.1 & 181209 & 1150(5) & 53472(6) & Neither \\
248933 & 142413.05+143904.0 & - & 2459(3) & 55008(4) & No Lines \\
249181 & 143426.77+093913.8 & - & 16288(11) & 71242(13) & No Lines, UL \\
249244 & 144022.48+082122.3 & - & 8966(4) & 62647(5) & No Lines \\
145944+102905 & 145944.89+102906.2 & - & 5884(3) & 59029(4) & No Lines \\
150338+121443$^7$ & 150342.94+121449.5 & 110830 & 2669(3) & 55255(4) & Neither, UL \\
HI150423.7+240930$^8$ & 150422.90+241004.6 & - & 1217(5) & 53550(6) & Pointing Error \\
HI150900 & 150900.54+085536.6 & 22768 & 16663(31) & 71682(36) & Neither, ML \\
258004 & 151236.29+151010.9 & - & 10552(4) & 64508(5) & No Lines, ML \\
257889 & 151546.67+155336.6 & 50026 & 11193(4) & 65261(5) & Neither \\
HI151620+152939 & 151620.26+152938.9 & 61857 & 12910(5) & 67277(6) & Neither, UL \\
151659+051751 & 151659.24+051751.5 & 15363 & 2221(13) & 54736(15) & Neither, ML \\
152933+150728 & 152933.22+150728.7 & 39468 & 5365(5) & 58420(6) & Neither, UL \\
727130 & 152948.19+260516.3 & - & 2019(3) & 54492(4) & No Lines \\
HI153050.1+123632 & 153052.73+123611.5 & 71904 & 629(4) & 52861(5) & Neither, UL \\
HI153948.8+275213 & 153950.39+275247.9 & - & 9322(5) & 63065(6) & No Lines, UL \\
\end{tabular}

\label{ambig}
\medskip
{$^1$}{The uncertainty on optical velocities is 130 km s$^{-1}$ for all objects.}

{$^2$}{Possible pointing error during observations; this object may not be the WISE bright source or the ALFALFA detection.}

{$^3$}{The $\alpha$.40 data release of the ALFALFA catalog incorrectly states that this object is an OHM \citep{alpha40}. While the measured optical velocity is much higher than the ALFALFA velocity, it does not match the OH velocity and the object's identity remains unknown.}

{$^4$}{The velocity determination for this object was measured from only one line, presumed to be H$\alpha$.}

{$^5$}{Bleed-in from a nearby star obscured optical lines for this object.}

{$^6$}{Two objects were observed within beam uncertainty of the ALFALFA detection. The first had no visible optical lines, and the second (the velocity listed in Table~\ref{ambig}) showed broad line emission that matched neither the OH nor the \ion{H}{i} velocity.}

{$^7$}{Due to high redshift, this object is likely an AGN.}

{$^8$}{Guiding errors during observing rendered these frames unusable; no further observations were made.}
\end{table*}

\begin{table*}
\footnotesize
\contcaption{Ambiguous optical counterparts.}
\begin{tabular}{cccccc}
\hline {Object Name} & \multicolumn{1}{|c|}{\parbox{2.1cm}{\centering Observed Position (J2000)}} & {Optical Velocity$^1$ (km s$^{-1}$)} & {\ion{H}{i} Velocity (km s$^{-1}$)} & {OH Velocity (km s$^{-1}$)} & {Designation}
\\ \hline
258212 & 154609.82+083131.0 & 40857 & 12576(6) & 66885(7) & Neither, ML \\
HI154718.3+043350 &154718.37+043346.7 & - & 5776(5) & 58902(6) & No Lines, UL \\
268216 & 161222.06+063217.5 & 82218 & 1723(3) & 54145(4) & Neither \\
268065 & 161643.74+151302.0 & 55184 & 10078(4) & 63952(5) & Neither \\
HI215549.4+303121 & 215551.26+303056.1 & - & 24(1) & 52151(1) & No Lines, UL \\
330051 & 230619.72+275337.8 & - & 6901(3) & 60223(4) & No Lines \\
HI231619+253530 & 231616.19+253534.5 & - & -1133(8) & 50792(9) & No Lines \\
333281 & 232230.19+243525.0 & - & 9708(2) & 63518(2) & No Lines; Star? \\
333335$^4$ & 233419.25+250753.1 & 40996 & 5112(11) & 58123(13) & Neither, UL \\
333476$^9$ & 235925.00+302828.4 & 50343 & 12076(5) & 66298(6) & Neither \\
\end{tabular}

\medskip
{$^4$}{The velocity determination for this object was measured from only one line, presumed to be H$\alpha$.}

{$^9$}{Two possible optical counterparts were observed. One matched neither OH nor \ion{H}{i} velocities, and the other had no clear optical lines. }
\end{table*}

\section{Analysis \& Discussion}
\subsection{ALFALFA OH Completeness}
For the survey to be `complete' with respect to OHMs, ALFALFA should have detected all previously known OHMs within the survey flux, area, and redshift limits; furthermore, the number of OHM detections should match predictions from the OH luminosity function \citep[OHLF,][]{OHLF}. Analysis of ALFALFA OH completeness provides a test of the OHLF in a blind survey, assesses the sensitivity of a blind \ion{H}{i} survey like ALFALFA to OHMs, and may provide guidance for future \ion{H}{i} surveys.

\citet{oh1} lists the majority of the $\sim$120 OH megamasers known prior to this work; only 8 of these previously known masers lie within the volume defined by the ALFALFA 40\% sky coverage \citep[listed in][]{alpha40} and depth ($0.167\le z_{OH} \le 0.244$). Maser spectra from the Arecibo Megamaser Survey show that only one of the eight previously known masers in the ALFALFA volume is above the survey's S/N detection limit of 4.6. This OHM, AGC 181310/IRAS 08201+2801, was discovered by \citet{oh2} at $z=0.1680$; it was indeed detected by ALFALFA (Table \ref{OHmatch}). The five new OHM detections presented in this work were not found in previous surveys because the objects did not yet have optical redshifts, which previous OHM surveys relied on for spectrometer tuning. While ALFALFA found all previously discovered OHMs within the constraints of the survey, it found only 12.5\% of the known OHMs within its sky footprint. This indicates that future \ion{H}{i} surveys must have lower detection threshholds than ALFALFA if detecting new OHMs is a secondary goal.

The OHLF \citep{OHLF} describes the expected power-law luminosity density distribution of OHMs:
\begin{equation}
\phi = (9.8 ^{+31.9} _{-7.5} \times 10^{-6}) L_{OH}^{-0.64 \pm 0.21} \text{Mpc}^{-3} \text{dex}^{-1}.
\label{OHLFgeneral}
\end{equation}
We note that the OHLF was developed from the results of a targeted survey, the Arecibo Megamaser Survey. The ALFALFA results provide the first opportunity to test the OHLF against the results of a blind survey.

To integrate the OHLF and find the number of expected OHMs, we must compute the volume of the 40\% ALFALFA data release as well as the survey's luminosity limits. Using the Cosmology Calculator \citep{ccalc}, the comoving volume within ALFALFA's redshift range is 3.17~Gpc$^3$. The survey covers $\sim$6.8\% of the sky, so we use 0.21 Gpc$^3$ as the total volume of ALFALFA. The survey does not have a hard upper luminosity cutoff; however, the OHLF was calculated using data below $10^{3.8} L_\odot$. For this analysis, we choose an upper luminosity limit of $10^4 L_\odot$. This is close enough to the bounds of the Arecibo Megamaser Survey that the OHLF should still be valid; furthermore, increasing the luminosity limit further will not dramatically change the number of additional OHMs expected due to the power-law luminosity drop off. ALFALFA's S/N detection limit of 4.6 corresponds to a 7.7 mJy peak in narrow Gaussian lines like OH; this determines the survey's lower luminosity limit. Assuming a Gaussian OH line with a line width of 150 km~s$^{-1}$ (typical for OHMs) and peak of 7.7~mJy, the integrated flux of the line is 0.87~Jy~km~s$^{-1}$. Assuming a distance corresponding to the maximum $z_{OH}\approx0.25$ for ALFALFA, the lower luminosity bound is $10^{3.2} L_\odot$. However, this is not a hard lower cutoff; narrow lines could be detected below our luminosity cutoff and broad lines could remain undetected above our luminosity cutoff. It is not surprising, then, that ALFALFA OHM 145537+062437 has a luminosity of 3.06 dex, below the 3.2 dex cutoff. There are also reliability limitations at the low-flux limit-- despite identifying candidate OHMs near $10^{3.2} L_\odot$, we could not measure optical redshifts for many of the faint optical counterparts and the objects remain in the `ambiguous' category (Table \ref{ambig}).

\begin{figure}
	\centering
	\includegraphics[width=.49\textwidth]{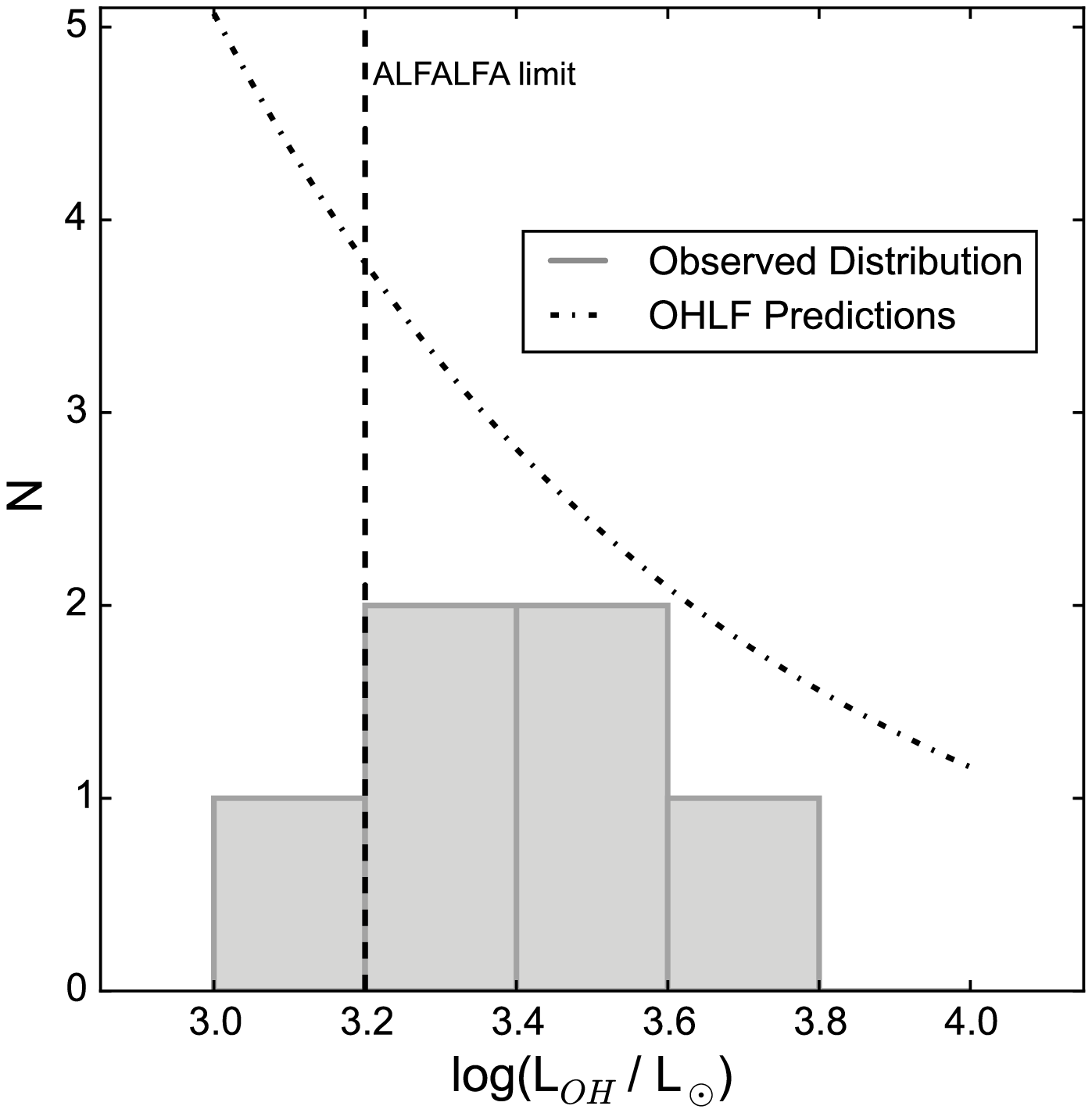}
	\caption{Number of OHMs as a function of luminosity. The histogram shows the observed ALFALFA OHM distribution; predictions from the OHLF are shown by the dashed line. Error bars on the OHLF curve are larger than the scale of the plot, and are not plotted.}
	\label{OHLF}
\end{figure}
Integrating Equation \ref{OHLFgeneral} from $10^{3.2} L_\odot$ to $10^{4.0} L_\odot$ over a volume of 0.21 Gpc$^3$ yields a prediction of $9_{-8}^{+73}$ OHMs in the ALFALFA $\alpha$.40 survey. We detected six OHMs in ALFALFA, well within the large error bars of our expectation. This indicates that OHMs found in a blind survey are consistent with the OHLF from the Arecibo Megamaser Survey. A plot of the OHLF and the distribution of observed OHMs is shown in Figure \ref{OHLF}.

ALFALFA found all previously known OHMs within the survey detection limits and the total number of OHMs in the survey is consistent with empirical predictions from the OHLF. We can therefore say that, within error, the ALFALFA survey is complete with respect to OHMs.


\subsection{Comparison of New and Existing OHMs}
Previously detected OHMs were primarily found through targeted surveys \citep[e.g.][Arecibo Megamaser Survey]{baan85, staveley, norris} that used IR selection criteria; however, this work makes use of a blind \ion{H}{i} survey to find OHM candidates and does not rely on assumed IR characteristics. Comparing ALFALFA OHMs with previously discovered masers thus provides an opportunity to verify the selection criteria used in previous targeted surveys and to test for a new OHM-producing environment at z$\sim$0.2. 

We use a two-sided  Kolmogorov-Smirnov (K-S) statistical test to determine if the two OHM populations come from the same distribution in IR space. The OHM comparison sample consists of 109 OHM host galaxies identified in \citet{baan} and the Arecibo Megamaser Survey. We use magnitudes and colors from the WISE All-Sky Source Catalog \citep{wise} as well as 60 and 100~$\mu$m flux from IRAS \citep{IRAS} and the FIR luminosity calculated from IRAS according to the prescription in \citet{FIRL}. IRAS did not detect ALFALFA OHMs AGC 219215 and 022657+282457, so they are not included in the K-S tests for IRAS 60 and 100~$\mu$m flux or FIR luminosity; all previously known OHMs are detected in IRAS. K-S test p-values are shown in Table~\ref{ks_newold}; all p-values are above 0.01, implying that the two populations do not show significant evidence for being drawn from different distributions. This indicates that the IR selection criteria used in previous targeted OHM surveys did not exclude a significant portion of the OHM population found in a blind survey; additionally, this work has not discovered a new population of OHMs missed by targeted surveys.

\begin{table}
\centering
\caption{K-S tests for ALFALFA OHMs and previously discovered OHMs.}
\begin{tabular}{cc}
\hline Infrared Property & K-S p-value  \\ \hline 
{WISE [3.4]} & 0.123 \\ 
{WISE [4.6]} & 0.326 \\ 
{WISE [12]} & 0.175 \\ 
{WISE [22]} & 0.103 \\ 
{WISE [12]} $-$ {[22]} & 0.038 \\ 
{WISE [4.6]} $-$ {[12]} & 0.797 \\ 
{WISE [3.4]} $-$ {[4.6]} & 0.174 \\ 
IRAS 60 $\mu$m flux & 0.055 \\ 
IRAS 100 $\mu$m flux & 0.015 \\ 
$\log(L_{FIR}/L_{\odot})$ & 0.055 \\
\hline
\end{tabular}
\label{ks_newold}
\end{table}


\subsection{Distinguishing OH from H {\normalsize \textbf{I}}}
\begin{figure}
\centering
\includegraphics[width=.49\textwidth]{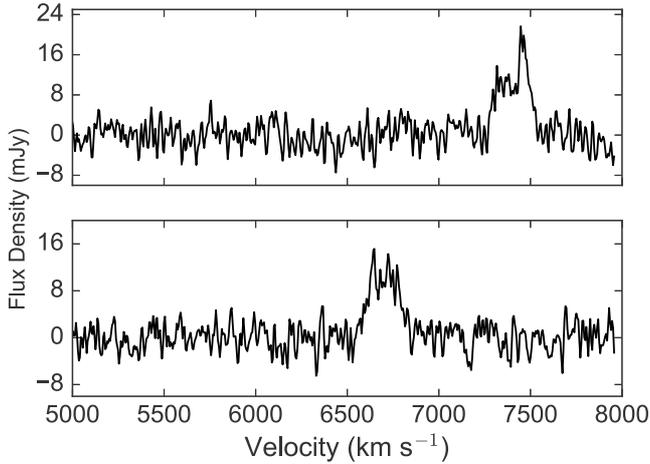}
\caption{Two sample ALFALFA spectra showing flux density as a function of \ion{H}{i} velocity \citep{alpha40}. Top is AGC 257959, an OH megamaser; bottom is AGC 230239, an \ion{H}{i} emission line galaxy.}
\label{spectra}
\end{figure}

It is impossible to distinguish OH lines from \ion{H}{i} lines using only \ion{H}{i} survey spectra. OHMs at z$\sim$0.2 have the same observed frequency as \ion{H}{i} at z$\sim$0.05, and the two radio lines are nearly indistinguishable. Example OH and \ion{H}{i} lines from ALFALFA are shown in Figure~\ref{spectra}. 

Given that it is difficult to distinguish between \ion{H}{i} and OH, how important is it for an \ion{H}{i} survey to separate the two populations? Only $\sim$0.05\% of ALFALFA $\alpha$.40 objects are OHMs. However, this is mostly due to the survey's low redshift range: the percentage of OH lines detected in an \ion{H}{i} survey is expected to increase with redshift because thermal \ion{H}{i} emission is mass-limited while OH maser amplification is not. Half of the detections from an \ion{H}{i} survey at a redshift of $z = 1$ are expected to be OH lines \citep{briggs}. 

Several high-redshift \ion{H}{i} surveys (including ASKAP-WALLABY \citep{askap}, MeerKAT-LADUMA \citep{meerkat}, and ultimately the Square Kilometer Array) are currently planned and in development. WALLABY will survey $z<0.26$, while LADUMA and SKA aim to cover redshifts up to z$\sim$1; all of these planned \ion{H}{i} surveys will observe a much larger population of OHMs than ALFALFA. Not only will locating OH lines in these surveys provide a larger and more robust sample of OHMs to further science goals such as tracking the galaxy merger rate as a function of redshift, it will also improve the fidelity of the survey catalogs. While OH lines in ALFALFA are not numerous, they are an excellent training set for distinguishing OH lines from \ion{H}{i} lines without the use of optical spectroscopy.

We expect OHMs, almost exclusively found in dusty IR-bright galaxies, to show different IR properties than \ion{H}{i} emitters. We used a K-S test in WISE colors and magnitudes to compare the 6 ALFALFA OHMs with ALFALFA \ion{H}{i} sources detected by WISE. We matched ALFALFA sources to the WISE All-Sky Source Catalog within a 45 arcsecond radius, requiring signal-to-noise $> 5$ at 3.4, 4.6, and 12~$\mu$m (bands W1, W2, and W3, respectively). Whenever optical counterparts were available, we used the optical coordinates for the match. WISE sources with the smallest offset from the ALFALFA coordinates were selected to be the IR counterpart. In total, 12,416 ALFALFA \ion{H}{i} sources were detected by WISE at 3.4, 4.6, and 12~$\mu$m. Only 5,801 of these \ion{H}{i} sources were detected by WISE at 22 $\mu$m, so only these objects are used for the [22] and [12] $-$ [22] K-S tests. The notation [22] refers to the Vega magnitude measured at 22 $\mu$m. Results of the K-S tests are tabulated in Table \ref{ks_hioh} and indicate that \ion{H}{i} and OH lines are significantly different in [4.6]~$-$~[12] and [3.4]~$-$~[4.6] colors.

\begin{table}
\centering
\caption{K-S tests for ALFALFA OHMs and ALFALFA \ion{H}{i} lines. Significant results are indicated in bold type, showing inconsistent [4.6]$-$[12] and [3.4]$-$[4.6]~$\mu$m colors between the two populations.}
\begin{tabular}{| c | c |}
\hline \multicolumn{1}{|c|}{{Infrared Property}} & \multicolumn{1}{c|}{{K-S p-value}}  \\ \hline 

{WISE [3.4]} & 0.072 \\ 
{WISE [4.6]} & 0.379 \\ 
{WISE [12]} & 0.018 \\ 
{WISE [22]} & 0.552 \\ 
{\textbf{{WISE [3.4]} $-$ {[4.6]}}} & $\boldsymbol{7.05 \times 10^{-6}}$  \\ 
{\textbf{{WISE [4.6]} $-$ {[12]}}} & \textbf{0.010} \\ 
{WISE [12]} $-$ {[22]} & 0.024 \\ 
\hline
\end{tabular}
\label{ks_hioh}
\end{table}

We made cuts in WISE color and magnitude space to separate \ion{H}{i} from OH line emitters. Cuts were chosen to include all OHMs and exclude as many \ion{H}{i} sources as possible; typically, the cut was made 0.1$-$0.2 dex above or below extremal OHM magnitudes or colors. Reducing the sample size was an iterative process: after making an initial cut and removing objects outside of the cut, we inspected the remaining sample and made a second cut in a different IR parameter. In total we made four cuts, listed in Table \ref{cuts}.  In addition to the two colors the K-S test indicates are inconsistent, we also made cuts in [3.4] and [22]. The [3.4] and [22] magnitude cuts were able to exclude a significant number of objects from the tails of the distributions. These objects did not significantly affect the p-values of the K-S test.

\begin{table}
\centering
\caption{Values for infrared cuts.}
\begin{tabular}{| c |}
\hline \multicolumn{1}{|c|}{{Infrared Cuts}}  \\ \hline 

[3.4] $-$ [4.6] $>$ 0.6 \\[0pt] 
[4.6] $-$ [12] $>$ 3.0 \\[0pt] 
[22] $>$ 4.8 \\[0pt] 
[3.4] $<$ 15.3 \\[0pt]
\hline
\end{tabular}
\label{cuts}
\end{table}

High values of [3.4] $-$ [4.6] and [4.6] $-$ [12] color select red objects, such as IR-luminous dusty ULIRGs likely to host OHMs. The 3.4~$\mu$m band corresponds to the stellar bump from old, red stars, so the [3.4] magnitude cut selects galaxies with large stellar mass (like ULIRGs) for a fixed redshift. OHMs are luminous at 22 $\mu$m, but are distant; local \ion{H}{i} sources could be brighter in [22] even though they are less luminous at 22 $\mu$m. So the high [22] cut removes nearby galaxies with low or moderate 22~$\mu$m luminosities. Cutting only bright 22~$\mu$m objects also avoids issues with incompleteness as it does not affect objects with low signal-to-noise in the WISE 22 $\mu$m band.

\begin{figure*}
\centering
\includegraphics[width=\textwidth]{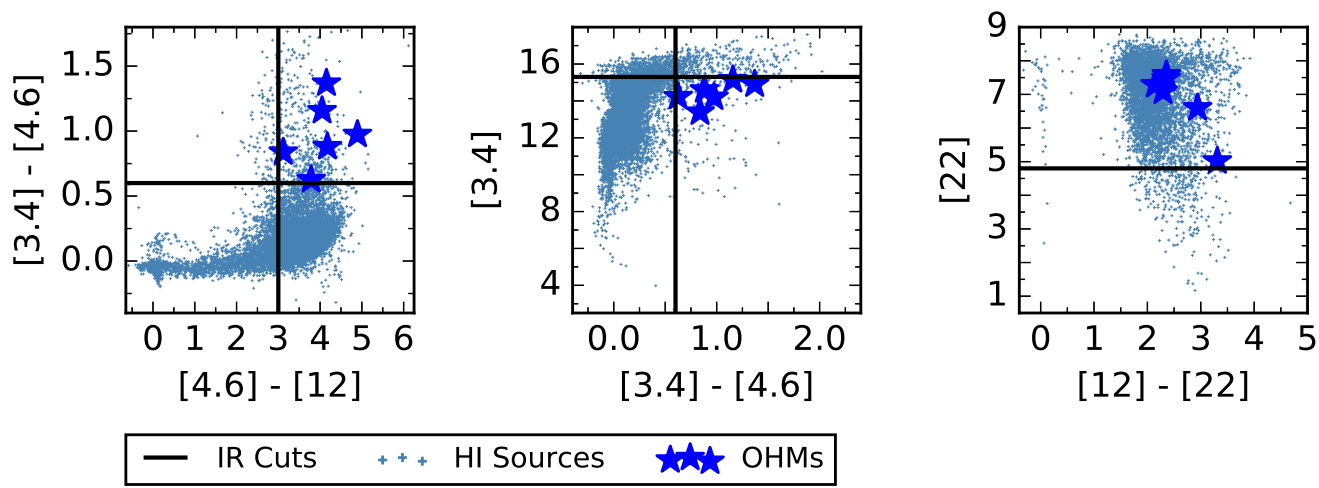}
\caption{The total ALFALFA \ion{H}{i} sample (blue crosses) along with ALFALFA OHMs (bright blue stars) in color-magnitude and color-color space. Black horizontal and vertical lines indicate the infrared cuts used to separate the samples (Table \ref{cuts}). (A color version of this figure is available in the online journal.)}
\label{cutcolor}
\end{figure*}

The total sample of ALFALFA \ion{H}{i} and OH sources as well as the infrared cuts from Table~\ref{cuts} are plotted in Figure~\ref{cutcolor}. While the [22] magnitude cut does not divide the sample to the same degree the other three cuts do, it removes $\sim$50 objects that the other cuts cannot distinguish from OHMs. We also see from Figure~\ref{cutcolor} that it is not necessary to modify our simple cuts into more complex cuts involving functions of more than one IR parameter-- slanted lines on the plots would not exclude a large number of additional objects. A final K-S test (Table \ref{postcut_KS}) on the post-cut samples confirms that further cuts in WISE space will not significantly improve the separation of OH and \ion{H}{i} line emitters.

\begin{table}
\centering
\caption{K-S tests for ALFALFA OHMs and \ion{H}{i} line emitters after imposing the IR color and magnitude cuts listed in Table~\ref{cuts}. The tests indicate that the OH- and \ion{H}{i}-emitting populations can no longer be distinguished in mid-IR color/magnitude space.}
\begin{tabular}{ccc}
\hline \multicolumn{1}{|c|}{{Infrared Property}} & \multicolumn{1}{c|}{{K-S p-value}} \\ \hline 

{[3.4]}  & 0.547 \\ 
{[4.6]}  & 0.136 \\ 
{[12]}  & 0.016 \\ 
{[22]}  & 0.748 \\ 
{[3.4]} $-$ {[4.6]} & 0.258 \\ 
{[4.6]} $-$ {[12]} & 0.038 \\ 
{[12]} $-$ {[22]} & 0.481 \\ 
\hline
\end{tabular}
\label{postcut_KS}
\end{table}

\begin{figure*}
\centering
\includegraphics[width=\textwidth]{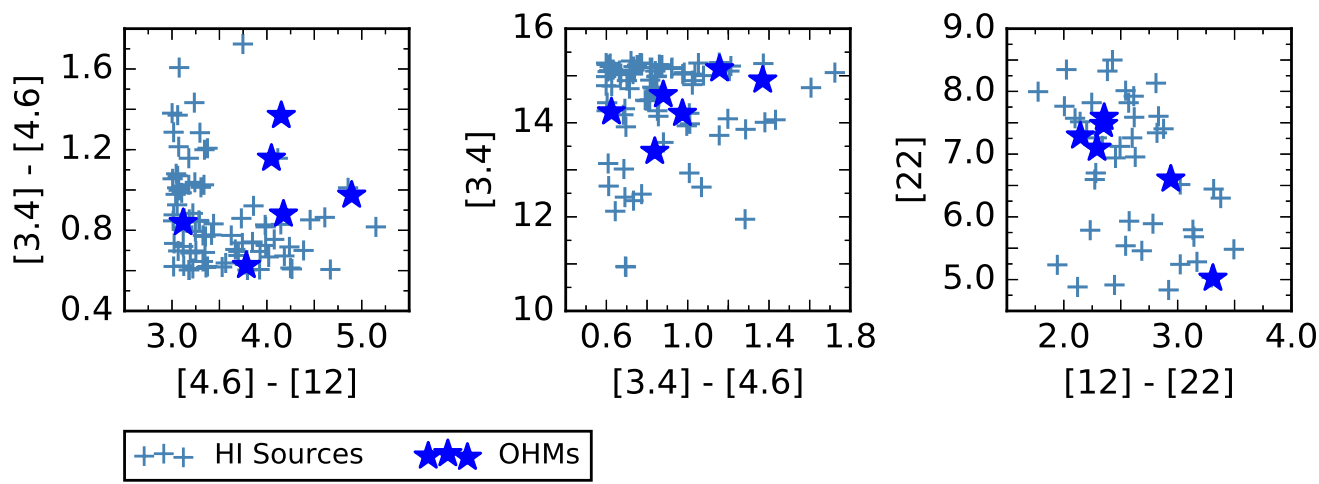}
\caption{The post-cut ALFALFA \ion{H}{i} sample (blue crosses) along with ALFALFA OHMs (bright blue stars) in WISE color-magnitude and color-color space. (A color version of this figure is available in the online journal.)}
\label{postcut}
\end{figure*}

ALFALFA objects that remain after the four IR cuts are plotted in color-color and color-magnitude space in Figure~\ref{postcut}. There are 83 remaining \ion{H}{i} sources, of which 43 are detected in WISE 22~$\mu$m. The IR cuts removed 99.3\% of all \ion{H}{i} sources and 99.3\% of \ion{H}{i} sources detected in WISE 22 $\mu$m, increasing the fraction of OHMs in the post-cut ALFALFA sample two orders of magnitude from $4.8\times 10^{-4}$ to $6.7\times 10^{-2}$. Other than a few \ion{H}{i} outliers, the OH and \ion{H}{i} detections that remain after the cuts are spread fairly uniformly across WISE space; this indicates that further cuts would not significantly improve the separation of the two populations. 

\subsection{ULIRG Redshift Evolution}
\begin{figure*}
\centering
\includegraphics[width=\textwidth]{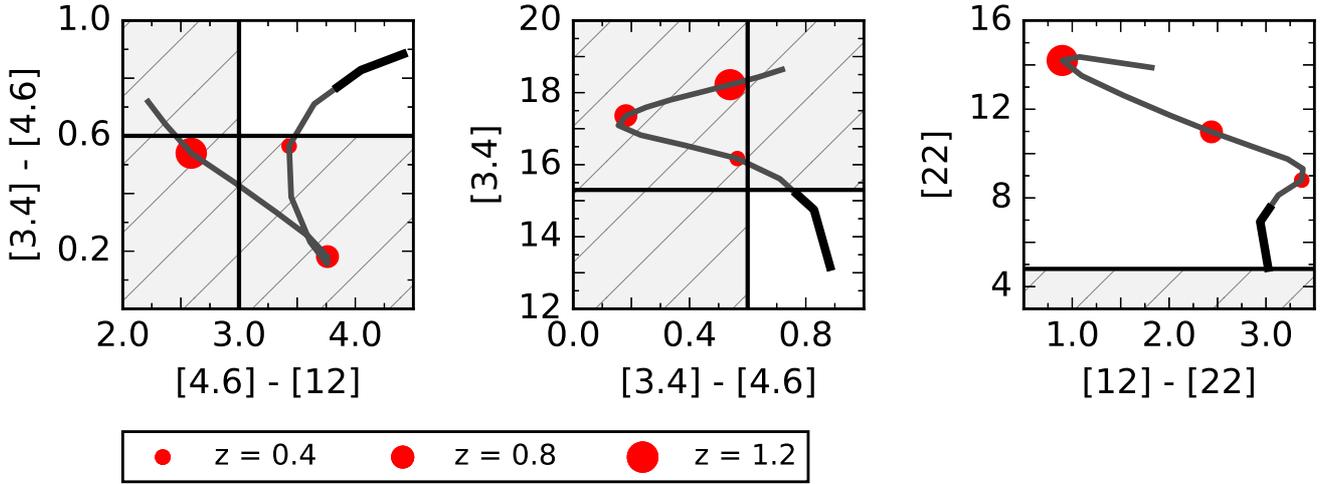}
\caption{Track of the redshift evolution of Arp 220. Red points of increasing area mark z~=~0.4, 0.8, and 1.2, while the bold portion of the track marks the ALFALFA redshift range. Horizontal and vertical black lines show the IR cuts, while the shaded region represents the quadrant or half of the plot that remains after the IR cuts. The tracks stay within the desired region through the ALFALFA redshift, z=0.1$-$0.25. Evolution past this point governs how the cuts should be adjusted for increasing redshift. (A color version of this figure is available in the online journal.)}
\label{zEvolution}
\end{figure*}

It is necessary to introduce a redshift dependence to the magnitude and color cuts listed in Table \ref{cuts} if the cuts are to be used for higher-redshift \ion{H}{i} surveys. We used Arp 220, the closest OHM and ULIRG, as a template spectral energy distribution (SED). The SED was created using Spitzer spectra from \citet{armus}; the optical and near-IR data were fit using two stellar components (old and young), and the far-IR SED was fit using dust continuum models from \citet{Chary}. In Figure \ref{zEvolution}, we plot the redshift evolution of Arp 220 and our infrared cuts in WISE color-color and color-magnitude space.

The Arp 220 tracks remain within the IR cuts for the ALFALFA OHM redshift range, $z_{OH}\le 0.25$. The tracks above this redshift indicate how the cuts should evolve for application to planned future \ion{H}{i} surveys. We note that the redshift evolution results in most OHMs being undetectable in the WISE 22 $\mu$m band by a redshift of $z\sim0.4$. However, it may be possible to continue to separate \ion{H}{i} and OH lines without use of the WISE 22 $\mu$m band or by applying the techniques used in this work to deeper IR catalogs.


\section{Conclusions}
In this work we identified OH megamasers in the ALFALFA \ion{H}{i} survey, confirmed their number and IR properties matched empirical predictions based on previous surveys, and developed a method to separate OH megamasers from 99\% of \ion{H}{i} line emitters without optical spectroscopy. Optical spectroscopy of 194 candidate OHMs in ALFALFA confirmed 129 uncertain \ion{H}{i} optical counterparts and discovered 5 new OHMs. 

Only one previously known OHM lay within the survey's volume and detection limits; this OHM was detected, raising the total number of OHMs in ALFALFA to six. Integrating the OHLF \citep{OHLF} over the sky footprint, redshift range, and luminosity detection limits of ALFALFA predicts $9_{-8}^{+73}$ OHMs in the survey, consistent with the six detections. 

Previous OHM surveys selected candidates based on IR properties; the ALFALFA OHMs are the first found in a blind spectral line survey. We used K-S tests to show that the two OHM populations come from the same statistical distribution in WISE color and magnitude as well as IRAS 60- and 100-$\mu$m flux and FIR luminosity. This validates the IR selection criteria used in previous surveys and suggests there is no previously unknown OHM-producing environment at $z\sim0.2$.

We used the WISE All-Sky Source Catalog to determine if OH line emitters can be distinguished from \ion{H}{i} line emitters without the use of optical spectroscopy. This is of particular interest for planned high-redshift \ion{H}{i} surveys, as the percentage of OH lines in an \ion{H}{i} survey is expected to increase with redshift and reach $\sim$50\% by $z=1$ \citep{briggs}. K-S tests confirm that the ALFALFA \ion{H}{i} and OH populations can be distinguished in WISE color and magnitude space, and four IR cuts (Table \ref{cuts}) reduce the total number of \ion{H}{i} objects from 12,416 to 83 and the number of \ion{H}{i} objects detected in WISE 22 $\mu$m from 5,801 to 43, excluding 99.3\% of the ALFALFA \ion{H}{i} sample. After the cuts, K-S tests could not distinguish the OH and \ion{H}{i} populations; this indicates further cuts in WISE space cannot further separate the two populations. It is possible that the OHM fraction could be increased using additional data such as galaxy morphology. The redshift evolution of the closest OHM, Arp 220, provides guidelines for adjusting the WISE color and magnitude cuts for higher-redshift \ion{H}{i} surveys. However, the sensitivity of the WISE 22~$\mu$m band precludes the use of a [22] $\mu$m cut at $z \gtrsim 0.4$. While a separation scheme based only on WISE data can distinguish OH from \ion{H}{i}, future work should investigate a separation scheme using additional data sources before the advent of high-redshift \ion{H}{i} surveys.

\section*{Acknowledgments}
{We thank the ALFALFA observing team as well as A. Truebenbach, who provided assistance calculating ULIRG redshift evolution. We thank the anonymous referee for helpful comments that substantially improved this manuscript. This work is based on observations made at the Apache Point Observatory, operated by New Mexico State University. Support for this work was provided by a grant from the Undergraduate Research Opportunities Program at the University of Colorado, Boulder and a CASA undergraduate research grant.  RG, MPH and the ALFALFA team at Cornell have been supported by NSF grants AST-0607007 and AST-1107390 and by grants from the Brinson Foundation. This research has made use of the NASA/IPAC Extragalactic Database (NED) which is operated by the Jet Propulsion Laboratory, California Institute of Technology, under contract with the National Aeronautics and Space Administration. This publication has made use of data products from the Wide-field Infrared Survey Explorer, which is a joint project of the University of California, Los Angeles, and the Jet Propulsion Laboratory/California Institute of Technology, funded by the National Aeronautics and Space Administration as well as data products from the Infrared Astronomical Satellite. This research also benefitted greatly from images provided by the Sloan Digital Sky Survey.}

\bsp	
\label{lastpage}

\begin{thebibliography}{}

\bibitem[Armus et al.(2004)]{armus} Armus, L., V. Charmandaris, H. W. W. Spoon, J. R. Houck, B. T. Soiffer, et al., 2004, \apjs, 154, 178

\bibitem[Baan, Wood, \& Haschick(1982)]{baanArp} Baan, W. A., P. A. D. Wood., A. D. Haschick, 1982, \apjl, 260, L49-L52

\bibitem[Baan, Haschick, \& Schmelz(1985)] {baan85} Baan, W. A., A. D. Haschick, J. T. Schmelz, 1985, \apjl, 298, L51-L54

\bibitem[Baan, Salzer \& LeWinter(1998)]{baan} Baan, W. A., J. J. Salzer, \& R. D. LeWinter, 1998, \apj, 509, 633-645

\bibitem[Briggs(1998)]{briggs} Briggs, F. H., \aap, 336, 815-822

\bibitem[Chary \& Elbaz(2001)]{Chary} Chary, R. \& D. Elbaz, 2001, \apj,556, 562-581

\bibitem[Clements et al.(1996)]{clements} Clements, Sutherland et al., 1996, \mnras, 277, 447-497, 1996.

\bibitem[Darling \& Giovanelli(2000)]{oh1} Darling, J. \& R. Giovanelli, 2000, \aj, 119, 3003-3014

\bibitem[Darling \& Giovanelli(2001)]{oh2} Darling, J. \& R. Giovanelli, 2001, \aj, 121, 1278-1293

\bibitem[Darling \& Giovanelli(2002a)]{oh3} Darling, J. \& R. Giovanelli, 2002a, \aj, 124, 100-126

\bibitem[Darling \& Giovanelli(2002b)]{OHLF} Darling, J. \& R. Giovanelli, 2002b, \aj, 572, 810-822

\bibitem[Darling(2007)]{gasfrac} Darling, J., 2007, \apjl, 669, L9-L12

\bibitem[Fullmer \& Lonsdale(1980)]{FIRL} Fullmer, L. \& C. Lonsdale, 1980, (Pasadena: JPL)

\bibitem[Giovanelli et al.(2005)]{ALFALFAprop} Giovanelli, R., M. Haynes, B. Kent, et al., 2005, \aj, 130, 2598-2612

\bibitem[Haynes et al.(2011)]{alpha40} Haynes, M. P., R. Giovanelli, A. M. Martin, K. M. Hess, A. Saintonge, et al., 2011, \aj, 142, 142-170

\bibitem[Hinshaw et al.(2013)]{wmap} Hinshaw, G. F., E. Komatsu, D.N. Spergel, C. L. Bennett, J. Dunkley, et al., 2013, \apjs, 208, 20B

\bibitem[Holwerda et al.(2011)]{meerkat} Holwerda, B. W., S.-L. Blyth, A. J. Baker, 2011, The Spectral Energy Distribution of Galaxies Proceedings IAU Symposium, 284

\bibitem[Johnston et al.(2008)]{askap} Johnston, S., R. Taylor, M. Bailes, N. Bartel, C. Baugh et al., Experimental Astronomy, 22, 3, 151-273

\bibitem[Norris, Gardner, \& Whiteoak(1998)]{norris} Norris, R. P., F. F. Gardner, J. B. Whiteoak, 1998, \mnras, 237, 673-681

\bibitem[Robishaw, Quataert, \& Heiles(2008)]{zeeman} Robishaw, T., E. Quataert, \& C. Heiles, 2008, \apj, 680, 981-998

\bibitem[Saintonge(2007)]{saintonge} Saintonge, A, 2007, \aj, 133, 2087-2096

\bibitem[Saunders et al.(2000)]{IRAS} Saunders, W., W. J. Sutherland, S. J. Maddox, O. Keeble, S. J. Oliver et al., 2000, \mnras, 317, 55-63

\bibitem[Stavely-Smith et al.(1992)]{staveley} Staveley-Smith, L., R. P. Norris, J. M. Chapman, D. A. Allen, J. B. Whiteoak, A. L. Roy. , 1992, \mnras,  258, 725-737

\bibitem[Wright(2006)]{ccalc} Wright, E. L., 2006, Publications of the Astronomical Society of the Pacific, 118, 1711-1715

\bibitem[Wright et al.(2010)]{wise} Wright, E. L., P. R. M. Eisenhardt, A. K. Mainzer, M. E. Ressler, R. M. Cutri, et al., 2010, \aj, 140, 1868-1881

\end{thebibliography}
\end{document}